\documentclass[final,authoryear,3p,times,twocolumn]{elsarticle}
\usepackage[latin2]{inputenc}
\usepackage{graphicx}
\usepackage{amssymb}
\usepackage{amsthm}
\usepackage{ulem}

\journal{Applied Radiation and Isotopes 105(2015)26}

\begin{document}

\begin{frontmatter}



\title{Extension of the energy range of experimental activation cross-sections data of deuteron induced nuclear reactions on indium up to 50 MeV}


\author[1]{F. T\'ark\'anyi}
\author[1]{F. Ditr\'oi\corref{*}}
\author[1]{S. Tak\'acs}
\author[2]{A. Hermanne}
\author[3]{A.V. Ignatyuk}

\cortext[*]{Corresponding author: ditroi@atomki.hu}

\address[1]{Institute for Nuclear Research, Hungarian Academy of Sciences (ATOMKI),  Debrecen, Hungary}
\address[2]{Cyclotron Laboratory, Vrije Universiteit Brussel (VUB), Brussels, Belgium}
\address[3]{Institute of Physics and Power Engineering (IPPE), Obninsk, Russia}

\begin{abstract}
The energy range of our earlier measured activation cross-sections data of longer-lived products of deuteron induced nuclear reactions on indium were extended from 40 MeV up to 50 MeV. The traditional stacked foil irradiation technique and non-destructive gamma spectrometry were used. No experimental data were found in literature for this higher energy range. Experimental cross-sections for the formation of the radionuclides $^{113,110}$Sn, $^{116m,115m,114m,113m,111,110g,109}$In and $^{115}$Cd  are reported in the 37-50 MeV energy range, for production of $^{110}$Sn and $^{110g,109}$In these are the first measurements ever. The experimental data were compared with the results of cross section calculations of the ALICE and EMPIRE nuclear model codes and of the TALYS1.6 nuclear model code as listed in the on-line library TENDL-2014.
\end{abstract}


\end{frontmatter}

% \linenumbers

\section{Introduction}
\label{1}
In the frame of our systematic study of charged particle induced activation, we earlier investigated the activation cross sections induced by deuterons on natural indium targets \citep{TF2011}. The main aim was then to study the production possibility of the $^{113/113m}$In medical generator up to the available 40 MeV deuteron energy at the CYRIC (Tohoku University) AVF cyclotron. The activation cross sections on indium are however of importance not only for the mentioned generator, but four other medically relevant radionuclides can be produced: $^{114m}$In (49.51 d), $^{113m}$In (99.476 min), $^{111}$In (2.8047 d) and $^{110m}$In (69.1 min). For the last three, apart from direct reactions also decay from parent $^{113}$Sn (115.09 d), $^{111}$Sn (35.3 min) and $^{110}$Sn (4.11 h) is contributing. Taking into account that no earlier data are available above 40 MeV incident deuteron energy and having the possibility to irradiate in the 50 MeV beam of the LLN Cyclone 90 cyclotron, we decided to extend the energy range of our previous study. Similar investigation is in progress on cross sections of proton induced reactions on indium up to 70 MeV by using different accelerators. The possible applications of the experimental data are discussed in our 2011 publication. Here we summarize the new experimental results, including only a short summary of the specific characteristics of experimental technique and the data evaluation method. The theoretical results calculated with ALICE-D code \citep{Dityuk} and the EMPIRE-D code  \citep{Herman} were taken from the previous work, while for the  calculation with the Talys code \citep{Koning2007} the latest data available in the TENDL-2014 library \citep{Koning2013}, based on the TALYS 1.6 version were used.

\section{Experiment and data evaluation}
\label{2}

For the cross section determination an activation method based on stacked foil irradiation technique and followed by $\gamma$-ray spectrometry were used. In order to avoid contamination of the stack by indium or its activation products in the case of accidental overheating, the In targets consisted of sandwiches of thin In foils backed by 50 $\mu$m Al and covered by a thin (6 $\mu$m) Al foil. The stack bombarded for 3600 s with a 50 MeV deuteron beam of 32 nA at Louvain la Neuve consisted of a sequence of Rh, Al, Al-In-Al, Pd, Al, Nb, Al foils (thicknesses are in Table 1) repeated 9 times. The deuteron beam was degraded from 50 MeV down to 32 MeV in the last In foil of the stack, assuring a good energy overlap with the 2011 study. Additional details on experiment, on beam parameters and on the data evaluation can be found in our earlier works made at same accelerator by using the same technique \citep{Hermanne}. The main experimental parameters and the methods of data evaluation for the present study are summarized in Table 1. The used decay data are collected in Table 2. The experimental data were measured relative to the $^{27}$Al(d,x)$^{24}$Na monitor reaction (see Fig. 1). The uncertainty of the experimental points was estimated by the common technique according to the ISO guide \citep{error}. The calculated experimental uncertainties are as follows: number of target nuclei including non-uniformity (5 \%), incident deuteron flux (7 \%), peak area including statistical errors of counts (0.1-20 \%), detector efficiency (5 \%), $\gamma$-ray abundance and branching ratio (3 \%). Except for a few data points the total uncertainty of 10-15 \% was obtained as positive square root of the quadratic sum of the individual uncertainty sources. Possible additional uncertainties due to non-linear effects of half-lives and cooling time were not taken into account. 

\begin{table*}[t]
\tiny
\caption{Main parameters of the experiment and the methods of data evaluations}
\begin{center}
\begin{tabular}{|p{0.8in}|p{1.6in}|p{1.1in}|p{1.8in}|}
\hline
\multicolumn{2}{|c|}{\textbf{Experiment}}  & \multicolumn{2}{|c|}{\textbf{Data evaluation}} \\
\hline
Incident particle & Deuteron & Gamma spectra evaluation & Genie 2000, 
\citep{Canberra}, Forgamma \citep{Szekely} \\
\hline
Method & Stacked foil & Determination of beam intensity & Faraday cup 
(preliminary)Fitted monitor reaction (final)\citep{TF1991} \\
\hline
Target stack and thicknesses & Rh (26 $\mu$m), Al(50 $\mu$m), Al(6 $\mu$m)-In(50 $\mu$m) 
-Al(50 $\mu$m), Pd(8 $\mu$m), Al(50 $\mu$m), Nb(10 $\mu$m), Al(50 $\mu$m) block Repeated 9 
times & Decay data & NUDAT 2.6 \citep{Nudat} \\
\hline
Number of target foils & 9x7 & Reaction Q-values & Q-value calculator 
\citep{Pritychenko}  \\
\hline
Accelerator & Cyclone 90 cyclotron of the Universit\'e Catholique in 
Louvain la Neuve (LLN) Belgium & Determination of beam energy &  
\citep{Andersen}  (preliminary)Fitted monitor reaction 
(final) \citep{error, TF2001}  \\
\hline
Primary energy & 50 MeV & Uncertainty of energy & Cumulative effects of 
possible uncertainties \\
\hline
Irradiation time & 60 min & Cross sections & Isotopic cross section \\
\hline
Beam current & 32 nA & Uncertainty of cross sections & Sum in quadrature 
of all individual contribution \\
\hline
Monitor reaction, $[$recommended values$]$ & $^{27}$Al(d,x)$^{24}$
Na reaction & Yield & Physical yield \citep{Bonardi} \\
\hline
Monitor target and thickness & $^{27}$Al 50+6 $\mu$m & Theory & 
ALICE-IPPE \citep{Dityuk}, ALICE-IPPE-D 
\citep{Ignatyuk2010} EMPIRE \citep{Herman}; EMPIRE-D \citep{Ignatyuk2011}; TALYS \citep{Koning2013, Koning2014} \\
\hline
detector & HPGe & & \\
\hline
$\gamma$-spectra measurements & 4 series & & \\
\hline
Cooling times (h) & 6.5-8.2\newline 45.2-49.8\newline 116.2-125.0\newline 557-582 & & \\
\hline
\end{tabular}

\end{center}
\end{table*}

\begin{table*}[t]
\tiny
\caption{Decay data of the investigated reaction products \citep{Nudat, Pritychenko}}
\begin{center}
%\begin{tabular}{|l|l|l|l|}
\begin{tabular}{|p{0.4in}|p{0.5in}|p{0.4in}|p{0.4in}|p{0.7in}|p{0.9in}|p{0.7in}|}
\hline
\textbf{Nuclide} & \textbf{Half-life} & \textbf{Decay method} &\textbf{E$_{\gamma}$(keV)} & \textbf{I$_{\gamma}$(\%)} & \textbf{Contributing 
reaction} & \textbf{Q-value(keV)} \\
\hline
$^{113m}$Sn\newline 7/2$^{+}$\newline 77.382 & 21.4 min & EC 8.9\newline  IT 91.1 & 77 & 
0.501 & $^{113}$In(d,2n)\newline $^{115}$In(d,4n) & -4044.5\newline -20357.66 \\
\hline
$^{113g}$Sn\newline 7/2$^{+}$ & 115.09 d & EC 100 & 255.134\newline 391.698 & 
2.11\newline 64.97 & $^{113}$In(d,2n)\newline $^{115}$In(d,4n) & -4044.5\newline -20357.66 
\\
\hline
$^{111}$Sn\newline 7/2$^{+}$ & 35.3 min & EC 69.8\newline  $\beta^{+}$ 30.25 & 
372.31\newline 761.97\newline 954.05\newline 1152.98\newline 1610.47 & 0.42\newline 1.48\newline 0.51\newline 2.7\newline 1.31 & $^{113}$
In(d,4n)\newline $^{115}$In(d,6n) & -22575.87\newline -38889.03 \\
\hline
$^{110}$Sn\newline 0$^{+}$ & 4.11 h & EC 100 & 280.462 & 100 & $^{113}$
In(d,5n)\newline $^{115}$In(d,7n) & -30744.7\newline -47057.9 \\
\hline
$^{109}$Sn\newline 5/2$^{+}$ & 18 min & EC 93.6\newline  $\beta^{+}$ 6.6 & 
649.8\newline 1099.2\newline 1321.3 & 28\newline 30\newline 11.9 & $^{113}$In(d,6n)\newline $^{115}$(d,8n) & 
-42027.11\newline -58340.26 \\
\hline
$^{116m1}$In\newline 5$^{+}$\newline 127.267 & 54.29 min & $\beta^-$ 100 & 1097.3 & 56.2 & 
$^{115}$In(d,p)\newline decay of $^{116m2}$In & 4560.154 \\
\hline
$^{115m}$In\newline 1/2$^{-}$\newline 336.24417 & 4.486 h & IT: 95.0\newline  $\beta^{-}$: 
5.0 & 336.24 & 45.8 & $^{115}$In(d,pn) & -2224.566 \\
\hline
$^{114m}$In\newline 5$^{+}$\newline 190.3682 & 49.51 d & EC 3.25\newline  IT 96.75 & 
190.27\newline 558.43\newline 725.24 & 15.56\newline  3.2\newline 3.2 & $^{113}$In(d,p)\newline $^{115}$
In(d,p2n) & 5049.324\newline -11263.836 \\
\hline
$^{113m}$In\newline 1/2$^{-}$\newline 391.691 & 99.476 min & IT 100 & 391.698 & 
64.94 & $^{113}$In(d,pn)\newline $^{115}$In(d,p3n) & -2224.566\newline -18537.734 
\\
\hline
$^{111}$In\newline 9/2$^{+}$ & 2.8047 d & $\varepsilon$: 100 & 171.28\newline 245.35 & 90.7\newline 94.1 
& $^{113}$In(d,p3n)\newline $^{115}$In(d,p5n)\newline decay of $^{111m}$In and 
$^{111}$Sn & -19342.13\newline -35655.3 \\
\hline
$^{110m}$In\newline 2$^{+}$\newline 62.08 & 69.1 min & $\beta^{+}$ 61.3\newline  EC 38.7 & 
657.75 & 97.74 & $^{113}$In(d,p4n)\newline $^{115}$In(d,p6n)\newline $^{110}$Sn 
decay & -29333.6\newline -45646.7 \\
\hline
$^{110g}$In\newline 7$^{+}$ & 4.92 h & $\beta^{+}$ 0.0081\newline  EC 99.9919 & 
641.68\newline 657.75\newline 707.40\newline 937.478\newline 997.16 & 26\newline 98\newline 29.5\newline 68.4\newline 10.5 & $^{113}$
In(d,p4n)\newline $^{115}$In(d,p6n) & -29333.6\newline -45646.7 \\
\hline
$^{109}$In & 4.167 h & EC 95.44\newline  $\beta^{+}$ 4.56 & 203.5\newline 426.2\newline 623.5 & 
73.5\newline 4.12\newline 5.5 & $^{113}$In(d,p5n)\newline $^{115}$In(d,p7n) & 
-37387.74\newline -53700.9 \\
\hline
$^{115g}$Cd\newline {1/2}$^{+}$ & 53.46 h & $\beta^-$ 100 & 581.87\newline 1109.76 & 33.1\newline 62.6 
& $^{115}$In(d,2p) & -2894.172 \\
\hline
\end{tabular}

\end{center}
\begin{flushleft}
\tiny{\noindent Abundances in $^{nat}$In: $^{113}$In 4.29 \%, $^{115}$In 95.71 \%. In case of clustered emission change the Q values by: pn$\rightarrow$d: +2.2 MeV, p2n$\rightarrow$t: +8.5 MeV
}
\end{flushleft}
\end{table*}

%\setcounter{table}{0}
%\begin{table*}[t]
%\tiny
%\caption{continued}
%\begin{center}

%\begin{flushleft}
%\tiny{\noindent 
%}
%\end{flushleft}

%\end{center}
%\end{table*}

\begin{figure}
\includegraphics[width=0.5\textwidth]{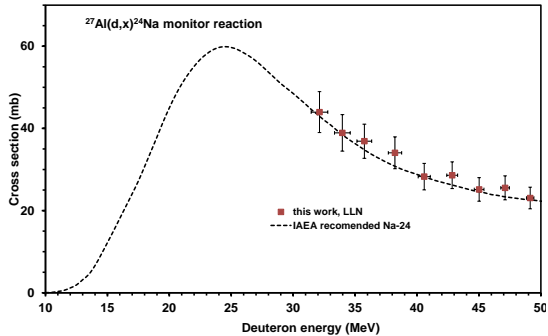}
\caption{Monitor reaction for $^{27}$Al(d,x)$^{24}$Na nuclear reaction}
\label{fig:1}       
\end{figure}

\section{Cross sections}
\label{3}
The measured cross sections for the production of $^{113,110}$Sn,$^{116m,115m,114m,113m,111,110g,109}$In and $^{115}$Cd are shown in Table 3 and Figures 2-11. The figures also show the theoretical results calculated with the ALICE-IPPE-D and the EMPIRE-D codes and the values available in the TALYS based TENDL-2014 library. The results for $^{115g}$Cd are significantly different compared what was published in our earlier work in the overlapping energy range. After a detailed check of both data evaluations an input mistake was discovered in the calculation sheet (the detector efficiency) of the earlier work. The results were corrected and are shown in Fig. 11. In most of the other excitation function a slight systematic shift can be observed, which could be caused by the different spectrometer system (efficiency calibration) and by the fact that all of our measurements are relative measurements to the monitor reactions. An improvement in the monitor reaction data can result in a modified excitation function by the nuclear reactions in question. The theoretical model calculations are taken from our previous work (except TENDL-2014 and for $^{110}$Sn, $^{110g}$In and $^{109}$In isotopes), which are new, compared to our previous work.

\subsection{$^{113}$Sn}
\label{3.1}
The excitation function of $^{113}$Sn (Fig. 2) was measured after total decay of the short-lived meta-stable state (21.4 min, IT 91 \%) to the long-lived ground state (115.1 d). An acceptable agreement with our earlier measurements can be seen. Clear improvement of description by the TALYS code in the new TENDL data can be observed.

\begin{figure}
\includegraphics[width=0.5\textwidth]{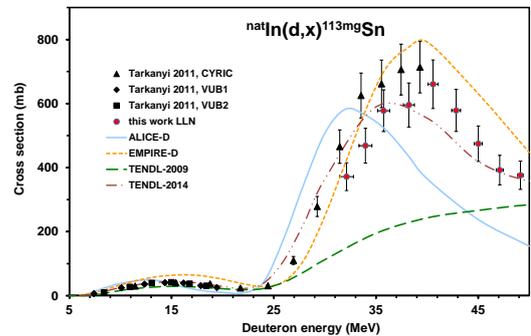}
\caption{Experimental and theoretical cross sections for the formation of $^{113}$Sn by the deuteron bombardment of indium}
\label{fig:2}       
\end{figure}

\subsection{$^{110}$Sn}
\label{3.2}
The experimental excitation functions for production of $^{110}$Sn (4.11 h) are shown in Fig. 3 in comparison with the TENDL-2014 predictions. Because it is a new series of data compared to or previous work in the same topic \citep{TF2011}, only the new TENDL-2014 predictions were re-calculated. No earlier experimental data exist.

\begin{figure}
\includegraphics[width=0.5\textwidth]{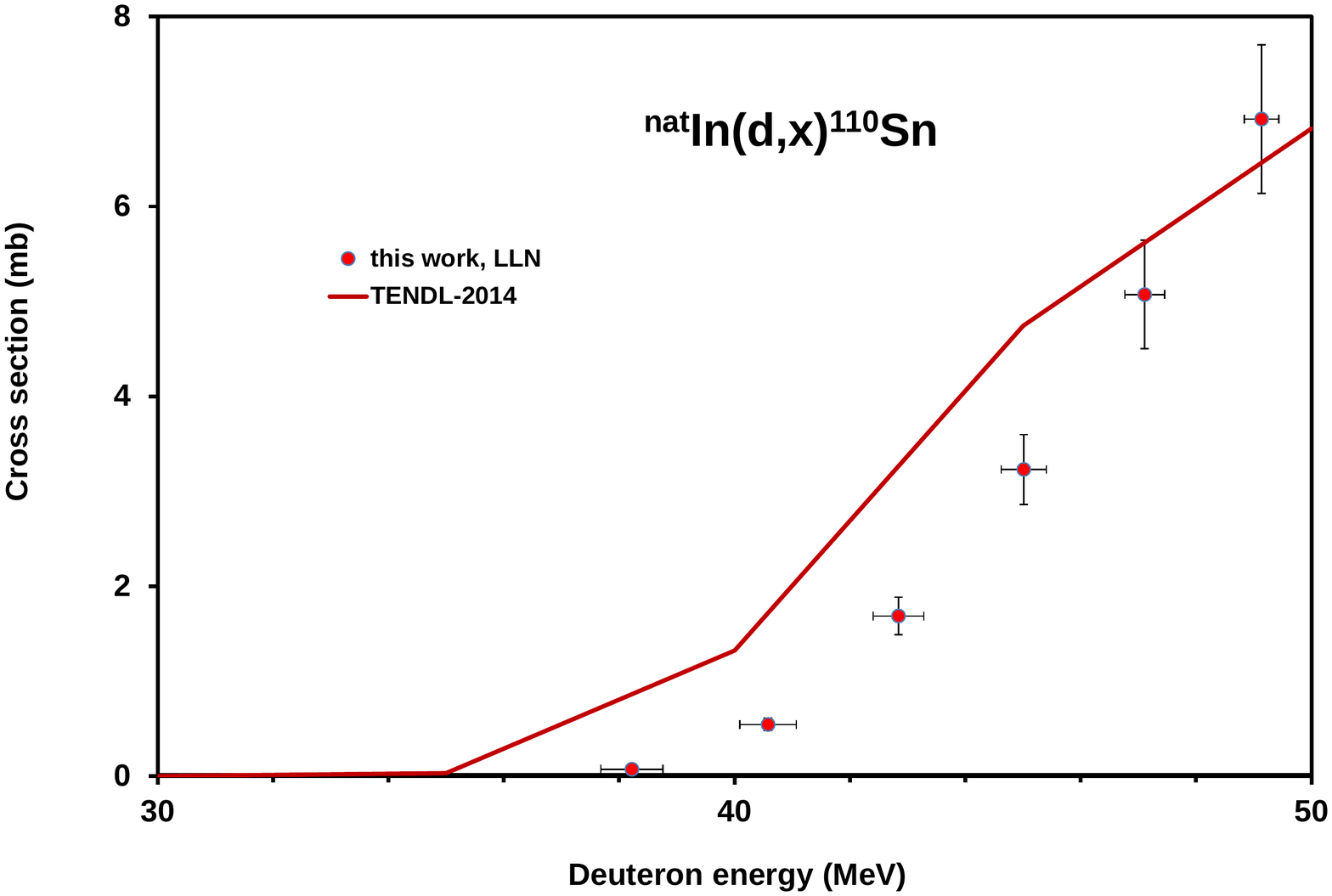}
\caption{Experimental and theoretical cross sections for the formation of $^{110}$Sn by the deuteron bombardment of indium}
\label{fig:3}       
\end{figure}

\subsection{$^{116m1}$In(m2+)}
\label{3.3}
The cross sections of the first metastable state (54.29 min)   were measured after complete decay of the second metastable state (2.18 s), which decays to m1 by 100 \% IT (Fig. 4). An acceptable agreement with our earlier measurements has been observed. No real change in description by TALYS code was done. All theoretical model calculations underestimate the experimental values except ALICE-D above 35 MeV.

\begin{figure}
\includegraphics[width=0.5\textwidth]{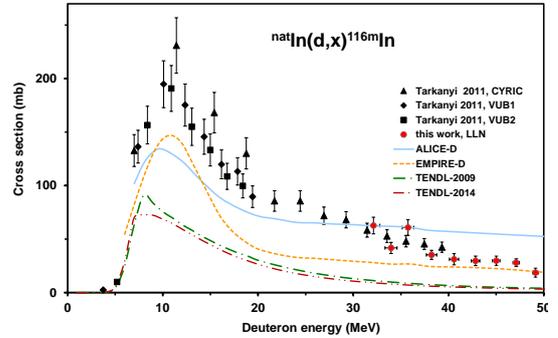}
\caption{Experimental and theoretical cross sections for the formation of $^{116m1}$In(m2+) by the deuteron bombardment of indium}
\label{fig:4}       
\end{figure}
\vspace{1 cm}

\subsection{$^{115m}$In}
\label{3.4}
The $^{115m}$In metastable state (4.486 h) can be produced directly via the $^{115}$In(d,pn) reaction and through decay of $^{115}$Cd. No gamma-lines from the decay of significantly longer-lived isomers of $^{115}$Cd were detected in the spectra measured shortly after EOB, indicating the negligible contribution to the $^{115m}$In production. The cross sections therefore can hence be considered as direct independent production cross sections (Fig. 5). An acceptable agreement with our earlier measurements is seen. The description by the recent version of the TALYS code is better than before.

\begin{figure}
\includegraphics[width=0.5\textwidth]{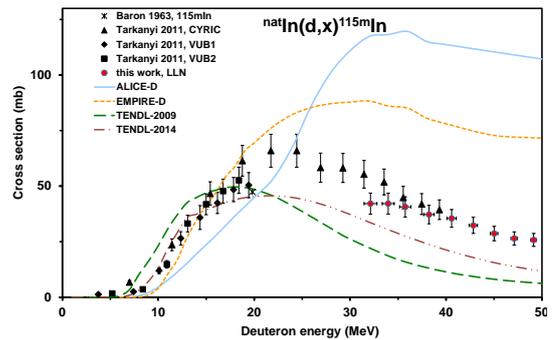}
\caption{Experimental and theoretical cross sections for the formation of $^{115m}$In by the deuteron bombardment of indium}
\label{fig:5}       
\end{figure}

\subsection{$^{114m}$In}
\label{3.5}
The independent cross sections for formation of the metastable state of the $^{114}$In ($T_{1/2}$ = 49.51 d) are shown in Fig. 6. The contribution of reactions on both stable indium isotopes can be distinguished. Our new results connect well to the earlier measurements, but give lower values in the overlapping energy range The description of the $^{113}$In(d,p)$^{114m}$In reaction is not improved in  the newest version of the TALYS code.

\begin{figure}
\includegraphics[width=0.5\textwidth]{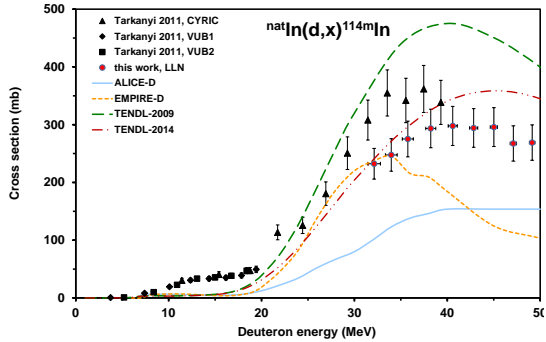}
\caption{Experimental and theoretical cross sections for the formation of $^{114m}$In by the deuteron bombardment of indium}
\label{fig:6}       
\end{figure}

\subsection{$^{113m}$In}
\label{3.6}
The $^{113m}$In (99.476 min) can be produced directly via (d,pxn) reaction and through decay of long-lived $^{113g}$Sn (115.09 d). Based on the measured $^{113}$Sn cross section (see above), no significant contribution of parent decay was estimated in the  first gamma spectra used for calculation of  $^{113m}$In production cross sections (Fig. 7). Good agreement with our earlier measurement and slight improvement of the description by the TALYS code were observed.

\begin{figure}
\includegraphics[width=0.5\textwidth]{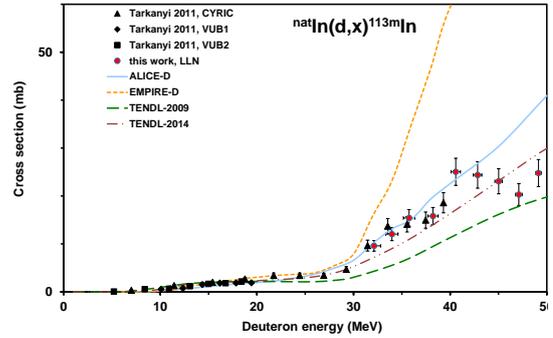}
\caption{Experimental and theoretical cross sections for the formation of $^{113m}$In by the deuteron bombardment of indium}
\label{fig:7}       
\end{figure}

\subsection{$^{111}$In}
\label{3.7}
The measured cross sections (Fig. 8) of $^{111g}$In (2.8047 d) are cumulative, including also production through  isomeric transition of the short-lived $^{111m}$In  metastable state (7.7 min) and decay of $^{111}$Sn parent (35.3 min). An acceptable agreement with our earlier measurements and clear improvement of the TALYS description above 25 MeV were observed.

\begin{figure}
\includegraphics[width=0.5\textwidth]{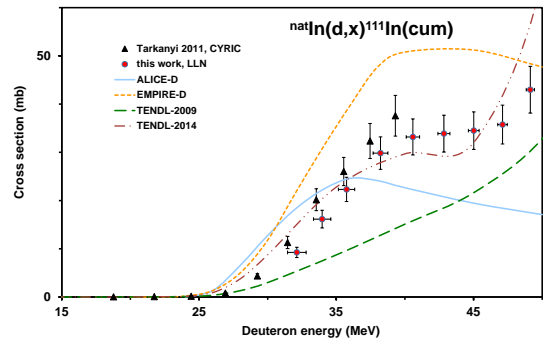}
\caption{Experimental and theoretical cross sections for the formation of $^{111}$In by the deuteron bombardment of indium}
\label{fig:8}       
\end{figure}

\subsection{$^{110g}$In}
\label{3.8}
The independent cross sections for production of $^{110g}$In (4.92 h) are shown in Fig. 9. No earlier experimental data were found. The description by TENDL-2014 is acceptable good, the other theoretical model codes are not used in this case.

\begin{figure}
\includegraphics[width=0.5\textwidth]{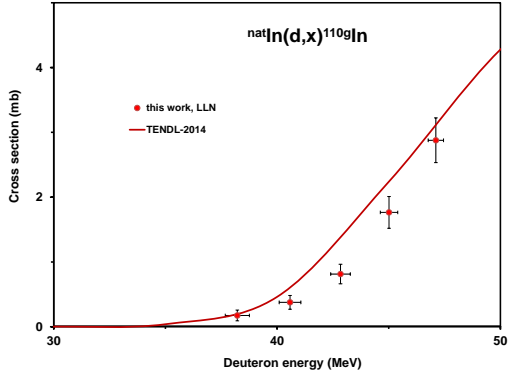}
\caption{Experimental and theoretical cross sections for the formation of $^{110g}$In by the deuteron bombardment of indium}
\label{fig:9}       
\end{figure}

\subsection{$^{109}$In}
\label{3.9}

We obtained cross sections for $^{109}$In (4.167 h) only at two high energy points (Fig. 10). No earlier data were found in the literature. The only theoretical model calculation with TENDL-2014 overestimates the experimental values.

\begin{figure}
\includegraphics[width=0.5\textwidth]{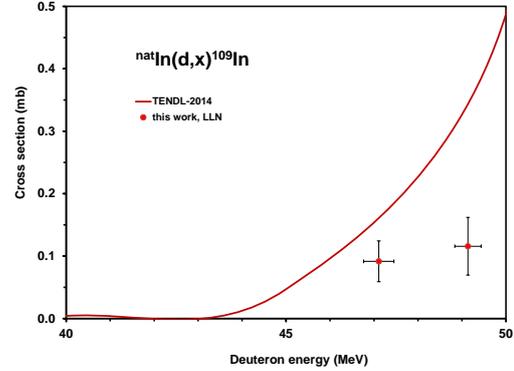}
\caption{Experimental and theoretical cross sections for the formation of $^{109}$In by the deuteron bombardment of indium}
\label{fig:10}       
\end{figure}

\subsection{$^{115g}$Cd}
\label{3.10}

The new independent cross section data  for the $^{nat}$In(d,x)$^{115g}$Cd process is  are shown in Fig. 11, together with the published and corrected data of \citep{TF2011} (see explanation of needed correction earlier) A good agreement exist between the two datasets. All codes give results that are about one order of magnitude off the experimental values.

\begin{figure}
\includegraphics[width=0.5\textwidth]{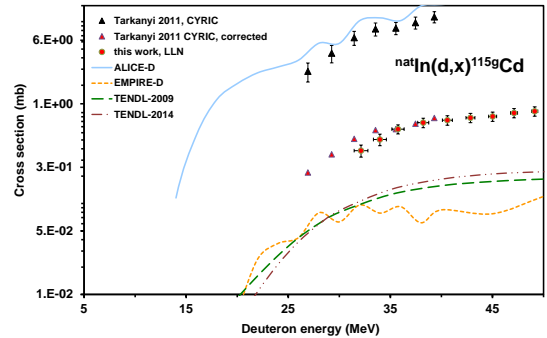}
\caption{Experimental and theoretical cross sections for the formation of $^{115m}$Cd by the deuteron bombardment of indium}
\label{fig:11}       
\end{figure}

\begin{table*}[t]
\tiny
\caption{Measured experimental data points for the $^{nat}$In(p,x) $^{113,110}$Sn, $^{116m,115m,114m,113m,111,110g,109}$In,$^{115}$Cd nuclear reactions}
\begin{center}
\begin{tabular}{|l|l|l|l|l|l|l|l|l|l|l|l|l|l|l|l|l|l|l|l|l|l|}
\hline
\multicolumn{2}{|c|}{\textbf{Energy}}& \multicolumn{20}{|c|}{\textbf{Cross section (mb)}} \\
\cline{3-22}
\multicolumn{2}{|c|}{(MeV)} & \multicolumn{2}{|c|}{\textbf{$^{113mg}$Sn}} & \multicolumn{2}{|c|}{\textbf{$^{110}$Sn}} & \multicolumn{2}{|c|}{\textbf{$^{116m}$In}} & \multicolumn{2}{|c|}{\textbf{$^{115m}$In}} & \multicolumn{2}{|c|}{\textbf{$^{114m}$In}} & \multicolumn{2}{|c|}{\textbf{$^{113m}$In}} & \multicolumn{2}{|c|}{\textbf{$^{111}$In}} & \multicolumn{2}{|c|}{\textbf{$^{110g}$In}} & \multicolumn{2}{|c|}{\textbf{$^{109}$In}} & \multicolumn{2}{|c|}{\textbf{$^{115g}$Cd}} \\
\hline
\textbf{E} & \textbf{$\Delta$E} & \textbf{$\sigma$} & \textbf{$\pm\Delta\sigma$} & \textbf{
$\sigma$} & \textbf{$\pm\Delta\sigma$} & \textbf{$\sigma$} & \textbf{$\pm\Delta\sigma$} & \textbf{$\sigma$} & 
\textbf{$\pm\Delta\sigma$} & \textbf{$\sigma$} & \textbf{$\pm\Delta\sigma$} & \textbf{$\sigma$} & \textbf{
$\pm\Delta\sigma$} & \textbf{$\sigma$} & \textbf{$\pm\Delta\sigma$} & \textbf{$\sigma$} & \textbf{$\pm\Delta\sigma$} & 
\textbf{$\sigma$} & \textbf{$\pm\Delta\sigma$} & \textbf{$\sigma$} & \textbf{$\pm\Delta\sigma$} \\
\hline
49.14 & 0.30 & 376.0 & 44.1 & 6.9 & 0.8 & 18.7 & 4.2 & 25.8 & 2.9 & 
269.0 & 30.6 & 24.8 & 2.8 & 43.0 & 4.8 & 4.1 & 0.5 & 0.12 & 0.05 & 1.03 
& 0.12 \\
\hline
47.11 & 0.34 & 392.3 & 46.1 & 5.1 & 0.6 & 28.1 & 3.5 & 26.5 & 3.0 & 
267.5 & 30.5 & 20.3 & 2.3 & 35.8 & 4.0 & 2.9 & 0.3 & 0.09 & 0.03 & 0.99 
& 0.12 \\
\hline
45.01 & 0.39 & 474.7 & 55.3 & 3.2 & 0.4 & 29.9 & 4.3 & 28.7 & 3.2 & 
295.8 & 33.6 & 23.1 & 2.6 & 34.5 & 3.9 & 1.8 & 0.2 &  & & 0.91 & 0.11 
\\
\hline
42.84 & 0.44 & 578.6 & 65.6 & 1.7 & 0.2 & 29.8 & 4.5 & 32.3 & 3.6 & 
294.3 & 33.5 & 24.4 & 2.8 & 33.9 & 3.8 & 0.81 & 0.15 &  & & 0.88 & 0.10 
\\
\hline
40.58 & 0.49 & 660.4 & 75.8 & 0.54 & 0.07 & 31.2 & 5.2 & 35.5 & 4.0 & 
297.8 & 33.9 & 25.0 & 2.8 & 33.2 & 3.7 & 0.38 & 0.11 &  & & 0.82 & 0.10 
\\
\hline
38.22 & 0.54 & 595.6 & 68.7 & 0.07 & 0.03 & 35.3 & 4.3 & 37.2 & 4.2 & 
293.4 & 33.4 & 15.8 & 1.8 & 29.8 & 3.4 & 0.17 & 0.08 &  & & 0.77 & 0.09 
\\
\hline
35.75 & 0.59 & 577.8 & 65.3 &  & & 60.9 & 7.3 & 40.7 & 4.6 & 275.2 & 
31.0 & 15.4 & 1.7 & 22.3 & 2.5 &  & &  & & 0.66 & 0.08 \\
\hline
33.97 & 0.63 & 468.5 & 54.4 &  & & 41.8 & 5.2 & 42.1 & 4.7 & 247.6 & 
28.2 & 12.1 & 1.4 & 16.2 & 1.8 &  & &  & & 0.51 & 0.07 \\
\hline
32.12 & 0.67 & 371.7 & 43.4 &  & & 62.8 & 7.8 & 42.1 & 4.7 & 232.4 & 
26.6 & 9.6 & 1.1 & 9.2 & 1.1 &  & &  & & 0.38 & 0.06 \\
\hline
\end{tabular}

\end{center}
\end{table*}

\section{Summary and conclusion}
\label{4}
We report experimental cross sections for production of $^{113,110}$Sn, $^{116m,115m,114m,113m,111,110g,109}$In and $^{115}$Cd in the 37-50 MeV energy range. The new data are first data sets for all products above 40 MeV and for production of $^{110}$Sn and $^{110g,109}$In no earlier experimental data were found. The experimental and theoretical data and the deduced integral yields were compared in detail in our 2011 publication. The comparison between the 2009 and 2014 versions of the TENDL library (obtained with the most recent version of the TALYS codes) shows further improvement for (d,xn) and some (d,pxn) reactions but the agreement is still poor when the (d,p) reactions plays an important role.
The possible use of experimental data for production was discussed in detail in our previous work and for $^{110m}$In, $^{111}$In, $^{113m}$In and $^{114m}$In will be included in our simultaneously submitted work on proton induced nuclear reactions on indium. The direct productions of indium radionuclides on indium target will in all cases be carrier added. Therefore for production of these medical radioisotopes via deuteron induced reactions on indium only the production possibility through generator parent isotopes should be taken into account.

\section{Acknowledgements}
\label{}
This work was performed in the frame of the HAS-FWO Vlaanderen (Hungary-Belgium) project. The authors acknowledge the support of the research project and of the respective institutions. We thank to Cyclotron Laboratory of the Universit\'e Catholique in Louvain la Neuve (LLN) providing the beam time and the crew of the LLN Cyclone 90 cyclotron for performing the irradiations. 
%\FloatBarrier
 


%% References

%\clearpage
\bibliographystyle{elsarticle-harv}
\bibliography{Ind}






% \end{thebibliography}

\end{document}